\documentclass[preprints,article,accept,moreauthors,pdftex]{Definitions/mdpi} 
\usepackage{amsfonts}
\usepackage{amssymb}
\usepackage{physics}
\usepackage{mathtools}
\usepackage{textcomp}
\usepackage{siunitx}
\usepackage{empheq}
\usepackage{bbold}
\usepackage{dcolumn}
\usepackage{bm}
\usepackage{wrapfig}
\usepackage{lipsum}

\firstpage{1} 
\pubvolume{1}
\issuenum{1}
\articlenumber{0}
\pubyear{}
\copyrightyear{}
\datereceived{}   
\dateaccepted{} 
\datepublished{{date}} 
\hreflink{https://doi.org/} 

\pdfoutput=1

\Title{Causality in a qubit-based implementation of a quantum switch}
\TitleCitation{Causality in a qubit-based implementation of a quantum switch}
 
\Author{Carlos Sab\'in $^{1,}$*\orcidA{}}

\AuthorNames{Carlos Sabín}
\AuthorCitation{Sabín, C.}
\address{%
$^{1}$ \quad Departamento de F\'isica Te\'orica, Universidad Aut\'onoma de Madrid, 28049 Madrid, Spain}
\corres{Correspondence: carlos.sabin@uam.es}

\abstract{We introduce a qubit-based version of the quantum switch, consisting in a variation of the Fermi problem. Two qubits start in a superposition state where one qubit is excited and the other in the ground state, but it is undefined which is the excited qubit. Then, after some time, if a photon is detected, we know that it must have experienced an emission by one atom and then an absorption and re-emission by the other one, but the ordering of the emission events by both qubits is undefined. While it is tempting to refer to this scenario as one with indefinite causality or a superposition of causal orders, we show that there is still a precise notion of causality: the probability of excitation of each atom is totally independent of the other one when the times are short enough to prevent photon exchange.}

\keyword{quantum switch; quantum causality; superconducting circuits}

\begin{document}



\section{Introduction}
The notion of causality in quantum physics seems to be experiencing a revisitation during the last years, due to new theoretical and experimental results revolving around the ``quantum switch'' \cite{brukner, brukner2, carabela, procopioexp,rubinoexp,exp3,exp4,exp5,exp6}. A quantum switch is an interferometric setup where the optical path of a photon depends on the state of a control qubit. If the qubit is in state 0, the photon will follow a path which includes undergoing two linear-optical operations A and B, while if the state is 1 the path includes the operations in the opposite order, first B and then A. Now, if the control qubit is in a superposition state of 0 and 1, then the path is undefined and so the order of the operations A and B. This is typically referred to as ``indefinite causal order" or "superposition of causal orders". However, it has been noted \cite{fabiocosta} that a more accurate phrasing would be that the order is entangled with the external control qubit. Tracing out the qubit, the system is left in a classical mixture of probabilities rather that in a superposition, which seems to be impossible to get without the aid of the external system. Note also that the quantum switch does not violate ``causal inequalities'' \cite{branciard} and therefore invoking a non-classical causal order is not necessary to explain the results. Actually, this seems to be a general result within standard quantum mechanics \cite{cannot}.

Relativistic causality and Lorentz invariance are the cornerstones of quantum field theory \cite{weinberg, eberhardross}. Quantum mechanics is a low-energy, fixed-number approximation of quantum field theory. Although not explicitly relativistic by construction, quantum mechanics respects the basic principle of causality, being a non-signallng theory where superluminal propagation of physical magnitudes is forbidden \cite{ghirardi,peres}. 
A paradigmatic example of causality at the quantum level is the so- called Fermi problem \cite{fermi}. At $t=0$, a two-level neutral atom $A$ is in its excited state and a two-level neutral atom $B$ in its ground state, with no photons present. The atoms do not interact directly, but can emit and absorb photons -- and therefore, exchange them. If they are separated by a distance $r$, the total probabilities for each atom of being in a given state are completely independent of the other atom at times $t<r/c$, being $c$ the speed of light \cite{ingvasion,milonni, powerthiru,fermiyo}, and therefore is impossible to extract any information at superluminal rates out of this system. 
\begin{figure}[t!]\centering
\includegraphics[width=0.5\textwidth]{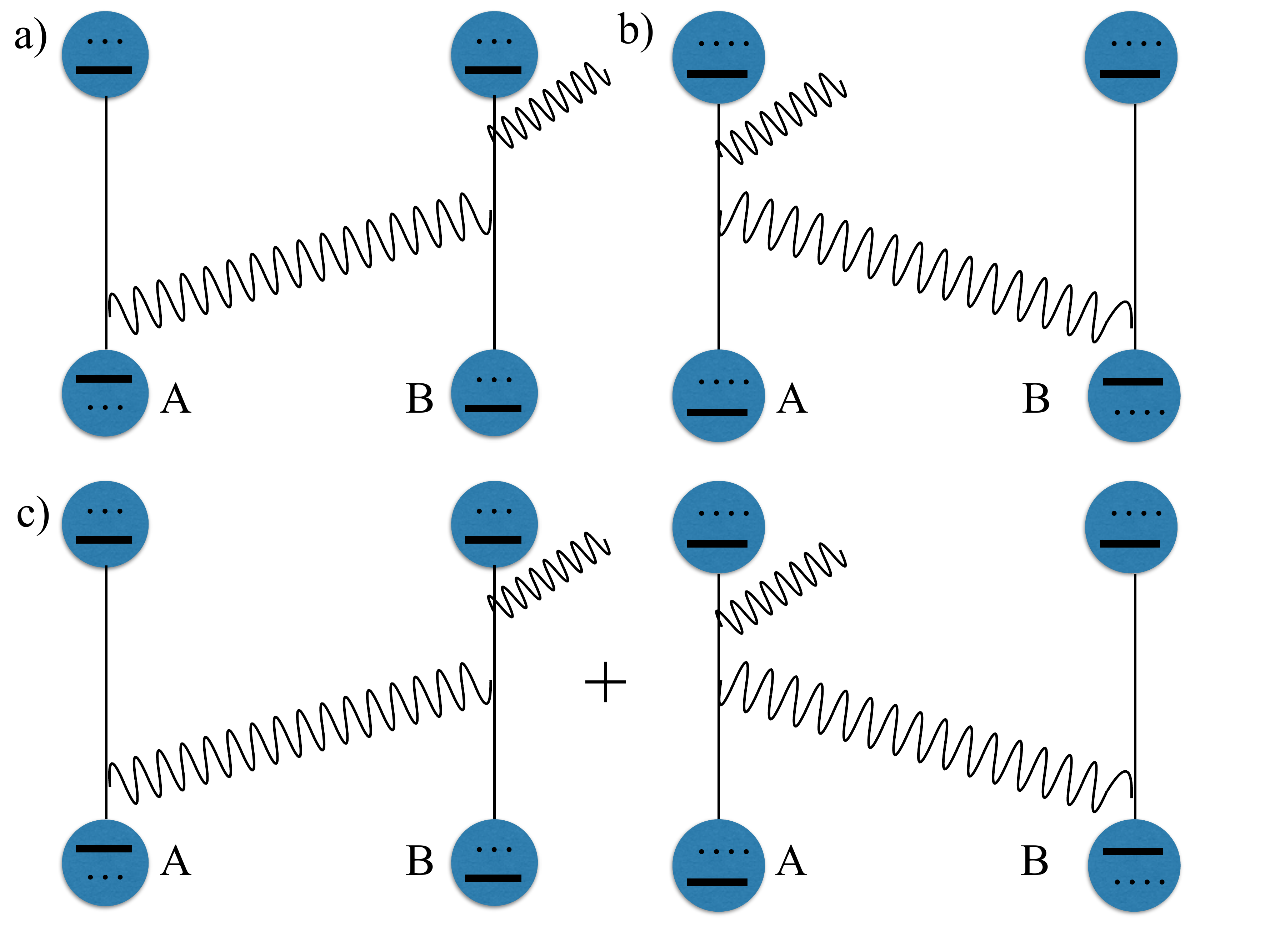}  
\caption{a) Qubit A is initially in the excited state while qubit B is initially in the ground state. A photon is emitted by qubit A and absorbed by qubit B, which then emits a photon and goes back to the ground state. b) Qubit B is initially in the excited state while qubit A is initially in the ground state. Qubit B emits a photon which is absorbed by qubit A and then qubit A emits a photon and relaxes back to the ground state. c) The initial state is a superposition of the initial states in scenarios a) and b). Then we know that a photon must be emitted, absorbed and re-emitted but the ordering of the emissions is not defined.}
   \label{Fig1}
  \end{figure} 

In this work we propose a modification of this model, which transforms the Fermi problem into a qubit-based version of the quantum switch. The atoms are connected to a beam-splitter through which we inject a single photon. Then, we still have one atom excited and the other in the ground state but it is undefined whether the excited atom is A or B. The atoms can emit and exchange photons. Under these conditions, if after some time a photon is detected, we don't know whether the atom has been emitted by A, absorbed by B and then re-emitted, or first emitted by B, then absorbed by A and re-emitted. Therefore, the ordering of the emission events is undefined and we have an analogue of the quantum switch. Note that A and B play the role of the local laboratories of Alice and Bob, which perform the quantum switch superoperation by exchanging the photon. The detected photon would play the role of the control qubit in the standard quantum switch. However, we will show that there is a precise notion of causality in this setup: the probabilities of excitation of the atoms do not depend on the other atom when they are outside the light cone of the other.

\section{Model and Results}


For simplicity, we focus
on a 1D setup with two qubits, $A$ and $B,$
interacting with a quantum bosonic field. The qubits are defined by two states
$\ket{e}$ and $\ket{g}$ with energy gap $\Omega$ ($\hbar=1$). They interact 
locally with a one-dimensional field, $E(x),$ 
\begin{eqnarray}
  E(x)=i\,\int dk \sqrt{N\omega_k}\,e^{ikx}a_k +\mathrm{H.c.}. \label{a}
\label{field}
\end{eqnarray}
This field would be a a 1D version of the electric field, mimicking the standard ``d times E" coupling in quantum optics. As usual, it is defined by a continuum of Fock operators
$[a_k,a^{\dag}_{k'}]=\delta_{kk'},$ and $\omega_k =
c|k|$, where $c$ is the speed of light. $N$ is a normalization constant which would depend on the particular setup.

We consider the standard Hamiltonian, $H =
H_0 + H_I,$ with the free part
\begin{equation}
  H_0 = \frac{1}{2}\Omega(\sigma^z_A + \sigma^z_B) + \sum_k \omega_k
  a^{\dagger}_ka_k, \label{b}
\end{equation}
and the interaction between them
\begin{equation}
  H_I =  d_A\,\sigma^x_A\, E(x_A)+d_B\, \sigma^x_B\,E(x_B). \label{c}
\end{equation}
Here $x_A$ and $x_B$ are the fixed positions of the atoms, and
$d_A$, $d_B$ are dipole moments.

The atoms do not interact directly, but they can emit and absorb photons, and therefore they can also exchange them. If the initial state contains a definite atom excited and the other in the ground state, for instance the state is $\ket{e_A\,g_B\,0}$, we have the classic Fermi model formulation (Fig. \ref{Fig1}a)-b)). In this 1D setup, it can be shown \cite{fermiyo} that the system behaves causally, meaning that the probability of excitation of atom B is independent of atom A for $t<r/c$, being $r=x_B-x_A$ the distance between the qubits. Note that this probability can be different from 0 for $t<r/c$, but this is not a causality problem as long as it is not dependent on qubit A and therefore cannot be used to transmit information. The probability of excitation of the initially unexcited qubit only becomes sensitive to the other qubit for $t>r/c$.

Now, let us consider that the atoms are connected to a 50-50 beam-splitter through which a single photon is injected. Then, if the qubits start in the ground state, after absorbing a photon they are initialized in the state:
\begin{equation}
  \ket{\psi} = \frac{1}{\sqrt2}(\ket{e_A\,g_B\,0}+\ket{g_A\,e_B\,0}), \label{d}
\end{equation}
in which one atom is excited, although we do not know whether is atom A or B, and no photons are present (Fig. (\ref{Fig1})c)). 
Then, after sufficiently long time, if a photon is detected we know that it has been emitted by one atom, absorbed by the other and reemitted. However, it is undefined whether it was first emitted by A, absorbed by B and re-emitted, or first emitted by B, absorbed by A and re-emitted. In this sense, we have built up a variant of the quantum switch: there are two events in the past of the photon -the two photo-emissions- whose temporal ordering is completely undefined. The detection of the photon plays the role of the quantum degree of freedom in the quantum switch: if photons are ignored the qubit system would be in a statistical mixture of both possibilities, not a truly quantum superposition. If time is not long enough to completely rule out the possibility of just a single emission, the source of the detected photon would still be indistinguishable and this would merely add new terms to the superposition, without changing the conclusions below.
 
The aim of this work is to show that, despite this indefiniteness in the temporal ordering, sometimes referred to as indefinite causality or superposition of causal orders, there is a well-defined notion of causality in this system. To this end, we are interested in the dynamics of the probability of excitation $P_{eA}$, $P_{eB}$ of atoms A and B. In the Heisenberg picture they would be given by:
\begin{eqnarray}
P_{ei}=\bra{\psi}\mathcal{P}^e_i (t)\ket{\psi}
\label{eq:probability}
\end{eqnarray}
where $ \mathcal{P}^e_i (t)$ is the projector over the excited state of atom $i$ 
\begin{equation}
\mathcal{P}^e_i=\ket{e_i}\bra{e_i}=\sigma^+_i\sigma^-_i 
\end{equation}
in the Heisenberg picture.
The Heisenberg equations are
\begin{equation}
\dot{\mathcal{P}}^{e}_{i} (t)=-i\, [\mathcal{P}^{e}_{i} (t), H]=- d_i\, \sigma^{y}_i (t) E(x_i,t),
\end{equation}
which give rise to
\begin{equation}
\mathcal{P}^{i}_{e} (t)- \mathcal{P}^{i}_{e} (0)=- d_i \int_0^t dt'\sigma^y_{i} (t') E(x_{i},t').
\label{eq:evolution}
\end{equation}

In this case, we have that the initial probabilities of excitation are 
\begin{equation}
\mathcal{P}^{A}_{e} (0)=\mathcal{P}^{B}_{e} (0)=1/2. 
\end{equation}
Therefore, putting together Eqs. (\ref{eq:evolution}) and (\ref{eq:probability}), we have that the probabilities of excitation would be given by:
\begin{equation}
P_{ei}=\frac{1}{2} -d_i \int_0^t dt' \bra{\psi}\sigma^y_{i} (t') E(x_{i},t')\ket{\psi}.\label{eq:pei}
\end{equation}
Let us now analyze $E(x_i,t)$.We start with  the Heisenberg equations of the operators $a_k$ and $a_k^{\dagger}$:
\begin{eqnarray}
\dot{a_k}(t) &=&-i\, [a_k (t), H]\nonumber\\&=&-i\omega_ka_k(t)-\sqrt{N\omega_k}\sum_i d_ie^{-ikx_i}\sigma_i^x(t)\nonumber\\
\dot{a^\dagger_k}(t) &=&-i\, [a^\dagger_k (t), H]\nonumber\\&=&i\omega_ka_k(t)-\sqrt{N\omega_k}\sum_i d_ie^{ikx_i}\sigma_i^x(t).
\end{eqnarray}
Integrating these equations, we get:
\begin{eqnarray}
a_k(t) &=&e^{-i\omega_kt}a_k(0)-\sqrt{N\omega_k}e^{-i\omega_kt}\sum_i d_ie^{-ikx_i}\int_0^t dt'\sigma_i^x(t')e^{i\omega_kt'}\nonumber\\
a^\dagger_k(t) &=&e^{i\omega_kt}a^\dagger_k(0)-\sqrt{N\omega_k}e^{i\omega_kt}\sum_i d_ie^{ikx_i}\int_0^t dt'\sigma_i^x(t')e^{-i\omega_kt'}.
\end{eqnarray}
Now, inserting them in Eq. (\ref{a}), the total field evaluated at  $x$ in Heisenberg picture is decomposed
\begin{equation}
E(x,t)= E_0(x,t)+E_A(x,t)+E_B(x, t)
\label{eq:fielddecomposed} 
\end{equation}
into the homogenous part of the field
\begin{equation}
E_0(x,t)=i\,\int_{-\infty}^{\infty}dk\,\sqrt{N\omega_k}\,e^{i(kx-\omega t)}a_k +\mathrm{H.c.}
\label{eq:fielddecomposed1} 
\end{equation}
and the back-action of $A$ and $B$ onto the field
\begin{eqnarray}
&&E_i(x,t)=-id_i\,N\times
\label{eq:fieldgeneral}\\
{}&&\quad\times \int_0^t \sigma_i^x (t')
\int_{-\infty}^{\infty}\omega_ke^{ik(x-x_i)-i\omega_k\,(t-t')}dkdt'
+\mathrm{H.c.}\nonumber
\end{eqnarray}
We now have to evaluate the field both in $x_A$ and $x_B$. Let us start  with $x_A$. Clearly, the field is decomposed into three parts:
\begin{equation}
E(x_A,t)= E_0(x_A,t)+E_A(x_A,t)+E_B(x_A, t).
\label{eq:fielddecomposedA} 
\end{equation}
Naturally, $E_0(x_A,t)$ cannot depend on atom B in any way. We should focus on $E_B(x_A,t)$ which carries an explicit dependence on atom B. We have:
\begin{eqnarray}
&&E_B(x_A,t)=-id_B\,N\times
\label{eq:fieldgeneral2}\\
{}&&\quad\times \int_0^t \sigma_B^x (t')
\int_{-\infty}^{\infty}\omega_ke^{ik(x_A-x_B)-i\omega_k\,(t-t')}dkdt'
+\mathrm{H.c.}\nonumber\\&&=d_B\,N\times\nonumber\\&& \frac{d}{dt}\int_0^t \sigma_B^x (t')
\int_{-\infty}^{\infty}e^{-ikr-i\omega_k\,(t-t')}dkdt'
+\mathrm{H.c.}\nonumber
\end{eqnarray}
Considering that $\omega_k=c|k|$, we have that the first integral is:
\begin{eqnarray}
&&\int_{-\infty}^{\infty}e^{-ikr-i\omega_k\,(t-t')}dk+\mathrm{H.c.}=\int_{0}^{\infty}e^{-ik(r+c\,(t-t'))}dk+\nonumber\\&&\int_{-\infty}^{0}e^{-ik(r-c\,(t-t'))}dk+\int_{0}^{\infty}e^{ik(r+c\,(t-t'))}dk+\nonumber\\&&\int_{-\infty}^{0}e^{ik(r-c\,(t-t'))}dk.
\end{eqnarray}
Performing the change of variables $k'=-k$ in the second and third integrals, this can be rearranged as:
\begin{eqnarray}
&&\int_{-\infty}^{\infty}e^{-ikr-i\omega_k\,(t-t')}dk+\mathrm{H.c.}=\int_{-\infty}^{\infty}e^{-ik(r+c\,(t-t'))}dk+\nonumber\\&&\int_{-\infty}^{\infty}e^{ik(r-c\,(t-t'))}dk.
\end{eqnarray}
The first integral is 0 --since $t'$ cannot be larger than $t$, by construction--, while the second gives rise to a Dirac delta:
\begin{equation}
\int_{-\infty}^{\infty}e^{ik(r-c\,(t-t'))}dk=2\pi\delta(ct'-(ct-r))=\frac{2\pi}{c}\delta\left(t'-(t-\frac{r}{c})\right).
\end{equation}
Plugging this result in Eq. (\ref{eq:fieldgeneral2}) we finally have that the B-dependent part of the field evaluated at $x_A$ is:
\begin{equation}
E_B(x_A,t)=\frac{2\pi d_B\,N}{c}\frac{d}{dt}\sigma_B^x (t-\frac{r}{c})\theta\left(t-\frac{r}{c}\right),\label{eq:eb}
\end{equation}
where the Heavyside step function makes explicit the fact that $E_B(x_A,t)$ is 0 for $t<r/c$.
We now turn to the A-dependent part of the field at $x_A$, which from Eq. (\ref{eq:fieldgeneral}) and following the same techniques as before leads to:
\begin{equation}
E_A(x_A,t)=\frac{2\pi d_A\,N}{c}\frac{d}{dt}\sigma_A^x (t).\label{eq:eA}
\end{equation}
This can be different from 0 for $t>0$. However, since $\sigma_A^x$ conmutes with the interaction hamiltonian, the commutator 
\begin{equation}\label{eq:comm1}
[H, \sigma_A^x]=i\Omega\sigma_A^y 
\end{equation}
does not involve atom B (see Eq. (\ref{eq:comm2}) below) and so do the commutators $[H, [H,\sigma_A^x]]$, $[H,[H, [H,\sigma_A^x]]]$ etc. Therefore the time-dependent operator $\sigma_A^x (t)$ cannot be dependent on atom B in any way.

Putting everything together we have shown that the time-dependent field evaluated at $x_A$, $E(x_A, t)$ is not dependent on atom B for $t<r/c$. There are B-independent parts that can be different from 0 at any time, but the B-dependent parts are only nonzero for $t>r/c$. Recalling Eq. (\ref{eq:pei}), we see that we still have to analyze the behavior of $\sigma_A^y(t)$. Again, the commutator 
\begin{equation}\label{eq:comm2}
[H, \sigma_A^y]=-i\Omega\sigma_A^x+2\,id_A\sigma_A^zE(x_A)
\end{equation}
does not contain any observable related with atom B and the same holds for all the commutators involved in the dynamics of $\sigma_A^y$. 

Therefore, we conclude that the probability of excitation of atom A $P_{eA}$ is completely independent of atom B for $t<r/c$. From Eq. (\ref{eq:pei}) we see that there will be three different terms $P_{eA0}$, $P_{eAA}$ and $P_{eAB}$ related with $E_0$, $E_A$ and $E_B$ respectively. The first two can be different from 0 at any time $t>0$, but are completely independent of atom B. The last one is only non-vanishing for $t>r/c$. 

In order to get more specific results, we can plug our exact results from Eqs. (\ref{eq:eb}) and (\ref{eq:eA}) into Eq. (\ref{eq:pei}) and retain only the leading terms in the commutator expansion of $\sigma_A^y(t')$ and $\sigma_A^x(t')$. Recalling Eqs. (\ref{eq:comm1}) and (\ref{eq:comm2}), we have:
\begin{eqnarray}
\sigma_A^y(t')&=&\sigma_A^y+ \mathcal{O} (\Omega t')\\
\sigma_B^x\left(t'-\frac{r}{c}\right)&=&\sigma_B^x-\Omega\left(t'-\frac{r}{c}\right)\sigma_B^y+\mathcal{O}\left((\Omega t')^2\right).\nonumber
\end{eqnarray} 

The, up to oder $d^2$ and $\Omega\,t$, we have:
\begin{eqnarray}\label{eq:peabpeaa}
P_{eAB}&=&\frac{2\pi d_Ad_B N \Omega}{c}\int_{\frac{r}{c}}^tdt'\bra{\psi}\sigma_A^y\sigma_B^y\ket{\psi}\nonumber\\&=&\frac{2\pi d_Ad_B N \Omega}{c}(t-\frac{r}{c})\theta\left(t-\frac{r}{c}\right)\\
P_{eAA}&=&\frac{2\pi d_A^2 N \Omega}{c}\int_0^t dt'\bra{\psi}\sigma_A^y\sigma_A^y\ket{\psi}\nonumber\\&=&\frac{2\pi d_A^2 N\Omega t}{c}\nonumber.
\end{eqnarray}
Interestingly, the first term only appears due to the fact that we are considering an initial superposition $\ket{\psi}$, since it comes from 
\begin{equation}
\bra{\psi}\sigma_A^y\sigma_B^y\ket{\psi}=1, 
\end{equation}
which would be 0 if we replace $\ket{\psi}$ with $\ket{e_A\,g_B\,0}$ or $\ket{g_A\,e_B\,0}$, as in the original Fermi problem. Thus, our modification of the Fermi setup in order to build a quantum switch does change the excitation probabilities of the atoms, but does not change the fact that they behave causally.

While all the analysis above refers to atom A, it also holds for atom B. Using again Eqs. (\ref{eq:fielddecomposed}), (\ref{eq:fielddecomposed1}) and (\ref{eq:fieldgeneral}) to evaluate the field at $x_B$, we have that the A-dependent part of the field at $x_B$ is: 
\begin{eqnarray}
&&E_A(x_B,t)=-id_A\,N\times
\label{eq:fieldgeneralB}\\
{}&&\quad\times \int_0^t \sigma_A^x (t')
\int_{-\infty}^{\infty}\omega_ke^{ik(x_B-x_A)-i\omega_k\,(t-t')}dkdt'
+\mathrm{H.c.},\nonumber
\end{eqnarray}
and it is straightforward to see that all the analysis above follows along the same lines.
Therefore, we have shown that the probabilities of excitation of both atoms are insensitive to the presence of the other atom until they enter the light cone of the other, as required by relativistic causality. These results apply in principle to any setup with two qubits interacting with a 1D bosonic quantum field --e. g. a superconducting-circuit setup such as in \cite{fermiyo}-- and could presumably be extended to 3D setups using similar Fermi-problem techniques \cite{milonni,powerthiru}. 

\section{Summary and Conclusions}

In summary, we have illustrated the relation between two notions that could seem to be in contradiction at first sight: on the one hand, the strong principle of causality underlying quantum theory and on the other hand, the recent developments around quantum switch devices, which seem to challenge causality with concepts such as superposition of causal orders or indefinite causality. To this end we have revisited a paradigmatic model of causality at the quantum level -- the two-atom Fermi problem-- and modified it in order to introduce a new version of the standard interferometric quantum switch, where the temporal ordering of two events --in this case, two photo-emissions by the atoms-- is not defined. We show that despite this indefiniteness in the ordering, the basic principle of causality survives, since the probabilities of excitation of both atoms behave in a causal way, as in the original Fermi problem. This suggests the necessity of careful distinctions \cite{fabiocosta,spekkens,paunk} between temporal ordering --which can be quantum-controlled and therefore indefinite-- and causal order, which is always well defined in standard quantum setups such as experimental quantum switches \cite{procopioexp,rubinoexp,exp3,exp4,exp5,exp6}. While modifications or extensions of quantum theory could lead to new causal structures that could be relevant in quantum gravity \cite{brukner2,zych,moller}, standard quantum mechanics does possess an standard causal structure.




\vspace{6pt} 




\funding{\textls[-25]{C.S. has received financial support through the Ram\'on y Cajal programme (RYC2019-028014-I)}.}

\end{paracol}
\reftitle{References}

\end{document}